\documentclass[aps, reprint, prapplied, twocolumn, superscriptaddress, amsmath, amssymb, longbibliography, noeprint]{revtex4-2}

\usepackage{times}
\usepackage{amsmath}
\usepackage{amssymb}
\usepackage{xfrac}
\usepackage{dsfont}
\usepackage{graphicx} 
\usepackage{dcolumn} 
\usepackage{bm} 
\usepackage{braket}
\usepackage{mathtools}
\usepackage{xcolor}

\newcommand{\affphys}{Department of Physics and Astronomy, Seoul National University, Seoul 08826, Korea}
\newcommand{\affiap}{Institute of Applied Physics, Seoul National University, Seoul 08826, Korea}
\newcommand{\affirc}{NextQuantum, Seoul National University, Seoul 08826, Korea}

\begin{document}

\title{Dual-species atomic absorption image reconstruction using deep neural networks}

% Author List
\author{Kyuhwan Lee}
\email{khlee3949@gmail.com}
\affiliation{\affphys}

\author{Yong-il Shin}
\email{yishin@snu.ac.kr}
\affiliation{\affphys}
\affiliation{\affiap}
\affiliation{\affirc}

%%%%%%%%%%%%%%%%%%%%%%%%%%%%%%%%%%%%%%%%%%%%%%%%%%%%%
% Abstract

\begin{abstract}
Optical imaging plays an instrumental role in understanding the behavior of trapped neutral atoms. 
In this work, we describe a deep learning-based online image completion protocol that reduces interference fringes in optical absorption signals for a dual-species atomic system. 
Regardless of the distinct nature of the task for two different atomic species, ${}^{6}$Li and ${}^{23}$Na, the method displays a robust solution for suppressing fringes. 
To incorporate this into daily operations, a transfer learning scheme is required that incrementally updates the previously learned parameters.
We outline an online image completion method that efficiently adapts to drifting experimental conditions.
Our method can be easily integrated into lab settings, where transfer learning can accelerate image analysis.
\end{abstract}

\maketitle

%%%%%%%%%%%%%%%%%%%%%%%%%%%%%%%%%%%%%%%%%%%%%%%%%%%%%
% Introduction
\section{Introduction}
Ultracold atomic systems provide a highly controlled environment for exploring quantum many-body dynamics.
Absorption imaging is one of the major probes for characterizing ultracold atomic systems. 
In the standard “double-shot” method, one image is taken \textit{with} atoms present and a second reference image is taken \textit{without} atoms; dividing the two yields the atomic absorption signal.
Analysis of the optical density (OD) profile from an in situ absorption image yields the atomic density distribution, while time-of-flight (ToF) imaging can be used to extract the momentum distribution of the atomic cloud \cite{ketterle1999making}.

A major issue in absorption imaging is the frequent appearance of fringe patterns in the processed image. 
These fringes arise from small differences in the probe light between the two exposures, typically caused by slight dislocations in the probe beam.
Since the probe beam itself often contains interference patterns, the mechanical vibrations during imaging would manifest as visible fringes in the processed image.
The fringe patterns constitute a significant source of noise, affecting the precision and accuracy of measured physical observables. 
Various techniques have been developed to mitigate fringes in absorption imaging. 
A direct approach is to improve the experimental stability, or to minimize the time delay between the signal and reference images.
This would relatively limit how much the probe beam can drift, thus reducing fringes.
A different approach would be to probe with incoherent light sources, which itself has a reduced interference fringe contrast \cite{sanner2012fluctuations, ma2014engineered, wei2019development}.

Beyond hardware solutions, post-processing algorithms have been proposed to reconstruct an ideal reference image to divide out the fringes. 
One class of methods uses principal component analysis (PCA) to form an ideal reference image using optimized weights of basis images \cite{erhard2004, kronjager2007, li2007reduction, ockeloen2010detection, niu2018optimized, song2020effective}.
Another class of methods relies on deep learning to comprehensively map the complex nonlinear relationship between the outer region without atoms and the inner region where atoms are located \cite{ness2020single, lei2022fringe}.  

In this work, we describe an extended implementation of a deep neural network (DNN) to flexibly address imaging artifacts under diverse experimental settings and drifting conditions.
We employ the dual-species configuration of our experimental setup, which offers a unique opportunity to investigate the extensive nature of the DNN in inferring reference images for each atomic species.
Preparation of a dual-species of ultracold atomic gases in a single integrated system has enabled a novel approach to studying the physics of quantum mixtures \cite{baroni2024quantum}.   
However, the different wavelengths that are used to probe the two atomic species may necessitate -- or present an advantage by allowing -- the two probe beams to take on different optical paths.  
We show that using a single trained DNN, a reference image can be extracted for both ultracold ${}^{6}$Li and ${}^{23}$Na gases that offer low fringe noise in the processed OD image, without specifying the atomic species in the image processing.
For PCA-based fringe removal, on the other hand, the two distinct optical paths may become a liability as the basis sets have to be individually prepared for the two different probe beams.  

Moreover, drifting experimental conditions, such as the alignment of the probe beam, require the DNN to be updated on a regular basis.
Rather than retraining a new network from scratch whenever conditions change, we leverage the initially trained DNN to fine-tune the optimized parameters with a small set of fresh calibration images. 
This allows the network to adapt in real time to the current state of the machine, maintaining high fringe-removal performance with a shorter training time. 
In the following, we present the design of our DNN and its training procedure, and demonstrate that this approach effectively suppresses fringe noise under a variety of experimental conditions.

%%%%%%%%%%%%%%%%%%%%%%%%%%%%%%%%%%%%%%%%%%%%%%%%%%%%%
% Method
\section{Method}
\subsection{Fringe-removal algorithm}

\begin{figure*}
\includegraphics{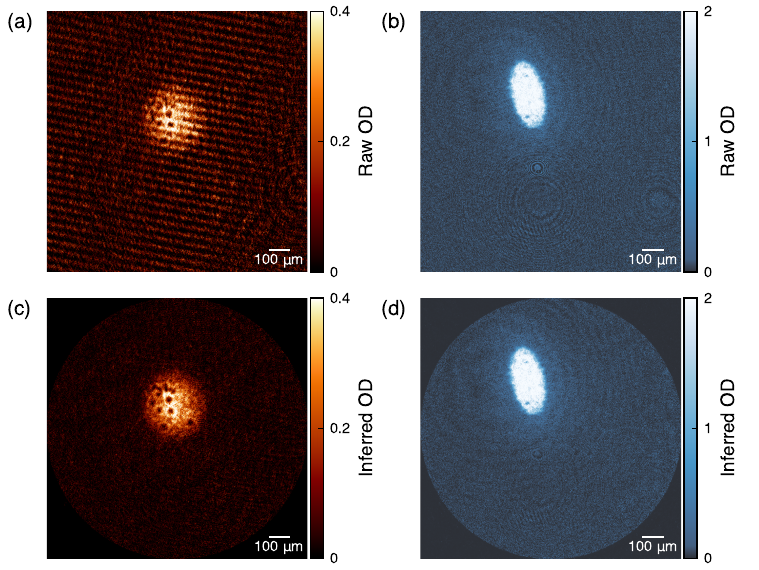}
\caption{
Comparison of optical density (OD) images before and after fringe removal using a deep neural network (DNN). 
Standard double-shot OD images for a unitary Fermi superfluid of ${}^{6}$Li (a) and a ${}^{23}$Na Bose-Einstein condensate (BEC) (b). The image in (a) was taken after a rapid ramp toward the BEC side of the BEC--Bardeen-Cooper-Schreiffer (BCS) crossover during a 22.5-ms time-of-flight (ToF), and in (b) after a 18-ms ToF, respectively.
(c, d) Corresponding OD images obtained using reference images inferred by the DNN.
The density depleted cores in the images represent quantum vortices.
Identical color scales are used for (a, c) and (b, d) for direct comparison. 
}
\end{figure*}

The main concept behind the removal of imaging artifacts is similar to \cite{ness2020single}, which relies on self-supervised learning.
We used reference images \textit{without} atoms as our ground truth.
After applying a circular mask to the reference images, we let our DNN predict the missing content within the mask based on the information outside the mask. 
For the loss function, we chose the root mean squared error (RMSE) of the predicted image with respect to the ground truth, which is minimized upon training. 
Applying this trained model to absorption images \textit{with} atoms, we can construct an atomic OD profile with significantly reduced imaging fringes [see Fig. 1 and Fig. 7 in the Appendix].

Our DNN architecture is based on a U-Net with skip connections that link the encoding and decoding layers \cite{ronneberger2015u} [see Fig. 6 in the Appendix].
We implemented our DNN using PyTorch.
The feedforward network comprises 38 convolutional layers and 7 max pool layers with a total of $0.5\times10^{9}$ trainable parameters.
Due to the limited memory of our GPU (NVIDIA RTX 4090), each batch contains only six images.
For the activation layers, we used a rectified linear unit (ReLU) function.
We used the ADAMW method \cite{loshchilov2017decoupled} with a weight decay of 0.0001 and He initialization \cite{he2015delving} for parameter optimization.

For normalization layers, we used group normalization.
Normalization layers generally accelerate the learning process \cite{ioffe2015batch, wu2018group}.
As suggested in \cite{wu2018group}, group normalization has proven to be an effective alternative to batch normalization for tasks such as high-resolution image reconstruction, where the large memory requirement limits the batch size. 
When transferring previously learned parameters to learn new tasks, the small batch size becomes a source of instability \cite{wu2018group}.
To implement a consistent transfer learning scheme, we used group normalization with the number of groups set to 16.

\subsection{Initial training details}

The DNN was trained on $768\times768$ pixel images. 
The circular mask had a radius of 384 pixels and was placed in the center of the image.  
Every image was normalized to have a value between 0 and 1.
The initial training dataset was prepared by selecting a subset of reference images acquired under various imaging conditions over a 12-week period, totaling 4370 images, with 20\% reserved as a validation set.

We ended the training sequence when the RMSE of the validation set has not improved by 0.2 \% over a patience value of 50 epochs.
The learning rate (LR) was set to $10^{-4}$ throughout the training process.
Initial training (or pre-training) took about 270 epochs or 60,000 s with NVIDIA RTX 4090, which was also identically used for post-training.
After the end of pre-training, the model exhibits an average RMSE of 0.033 on the entire pre-training dataset [See Fig. 2 and Appendix B].
Once the model is loaded, it takes less than 0.5 s to go through the feedforward network and reconstruct the OD image.
With the given hardware, for experimental sequences that take more than a second, the DNN can be easily integrated into the OD calculation process. 

%%%%%%%%%%%%%%%%%%%%%%%%%%%%%%%%%%%%%%%%%%%%%%%%%%%%%
% Method
\section{Results}

\subsection{DNN performance for a dual-species atomic system}

Our system uses magnetically trapped ${}^{23}$Na to sympathetically cool down ${}^{6}$Li to quantum degeneracy.
Details of our experimental apparatus on creating a strongly interacting Fermi gas of ultracold ${}^{6}$Li are outlined in \cite{park2018critical, lee2024universal}.
We leveraged the dual-species nature of our system to test whether the DNN can provide a robust solution for reconstructing ideal reference images for different atomic species.
The pre-training dataset comprises reference images from the standard double-shot method: for ${}^{23}$Na, we mostly used in situ and 5-ms to 18-ms ToF images; for ${}^{6}$Li, we mostly used in situ and 22.5-ms ToF images during which we rapidly ramp the Feshbach magnetic field toward the Bose–Einstein condensate (BEC) side of the BEC–Bardeen-Cooper-Schreiffer (BCS) crossover \cite{park2018critical, lee2024universal}.
For different ToF values, the camera position has to be adjusted to compensate for gravitational falling, resulting in qualitatively different reference images. 
The histogram of the RMSE for the entire training datatset after completing the initial training is shown in Fig.~2.
The concentrated distribution of the RMSE close to its average value indicates that the trained model does not display any significantly biased performance for different imaging conditions.

Figure 1 illustrates the performance of our model by comparing sample images of the standard double-shot approach [Figs. 1(a) and (b)] and model prediction [Figs. 1(c) and (d)].
Figs. 1(a) and (c) (Figs. 1(b) and (d)) show ToF images of a unitary Fermi superfluid gas of ${}^{6}$Li (a BEC of ${}^{23}$Na). 
Despite the qualitatively different background signal [Fig. 8 in the Appendix], the model effectively infers the reference images, thus significantly reducing fringes in the OD image. 
In particular, Fig. 1(a) and Fig. 1(c) showcase the practical utility of the deep learning-based fringe removal algorithm.
Quantum vortices, which reflect the underlying superfluidity of the degenerate gas, are represented by density-depleted cores in the ToF images.
In the standard double-shot method [Fig. 1(a)], however, quantum vortices are not clearly discernible.
However, upon reconstructing an ideal reference image using the trained model, the quantum vortices can be easily identified due to reduced contrast of the fringes [Fig. 1(c)].

To further shed light on the model performance and its versatility, we estimate the quality of the standard double-shot reference image in the experiment by calculating the RMSE between the first atom-shot image and the second reference image within a region where the number of atoms is negligible.
For the estimation, we selected 10 images for each different imaging conditions from the training dataset, resulting in a collection of 40 ${}^{23}$Na images and 50 ${}^{6}$Li images.
The RMSE calculation was performed for a $60\times60$ square patch located close to the edge of the circular mask, which barely contained any atoms.
The average RMSE within the patch between the first atom-shot image and the standard double-shot reference image for ${}^{6}$Li and ${}^{23}$Na images are 0.053 and 0.071, respectively~\footnote{In ${}^{6}$Li imaging, the average photon number is four times higher than that in ${}^{23}$Na imaging.}. 
Similarly, the average RMSE between the first atom-shot image and the inferred reference image within the same square patch for ${}^{6}$Li and ${}^{23}$Na images are 0.032 and 0.041, respectively.
The lower RMSE obtained by using inferred reference images compared to standard double-shot reference images clearly demonstrates the fringe-removal performance of our trained DNN.
Furthermore, the similar RMSE for different atomic species indicates that regardless of the qualitatively different probe beam profiles, the model can robustly infer the reference image for a dual-species atomic system.

\begin{figure}
\includegraphics{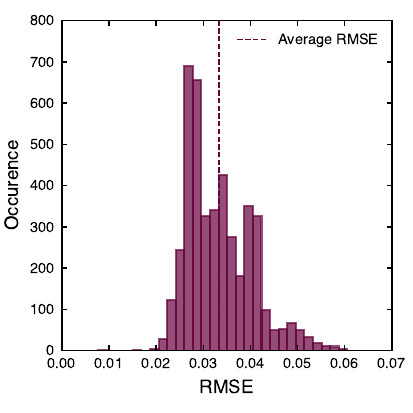}
\caption{
Histogram of the root mean squared error (RMSE) between inferred and ground truth reference images, across the entire initial training (or pre-training) dataset. 
The dashed vertical line indicates the average RMSE.
}
\end{figure}

\subsection{Transfer learning}

\begin{figure}
\includegraphics{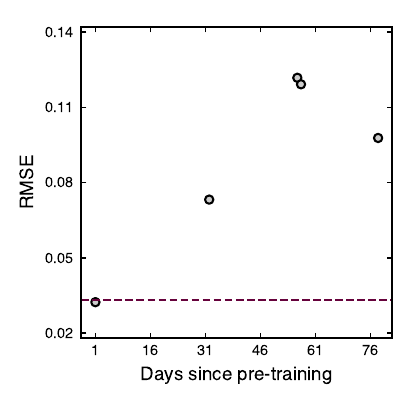}
\caption{
Deterioration of model performance for fresh data as a function of the number of days since pre-training. 
Each datapoint shows the average RMSE between ground truth reference images and reference images inferred with the pre-trained model.
The RMSE increases steadily over time, indicating that the performance of the pre-trained DNN in reconstructing ideal reference images degrades under drifting experimental conditions. 
The dashed line indicates the average RMSE of the pre-trained DNN on the pre-training data.
Each datapoint contains between 54 and 226 images.
}
\end{figure}

\begin{figure*}
\includegraphics{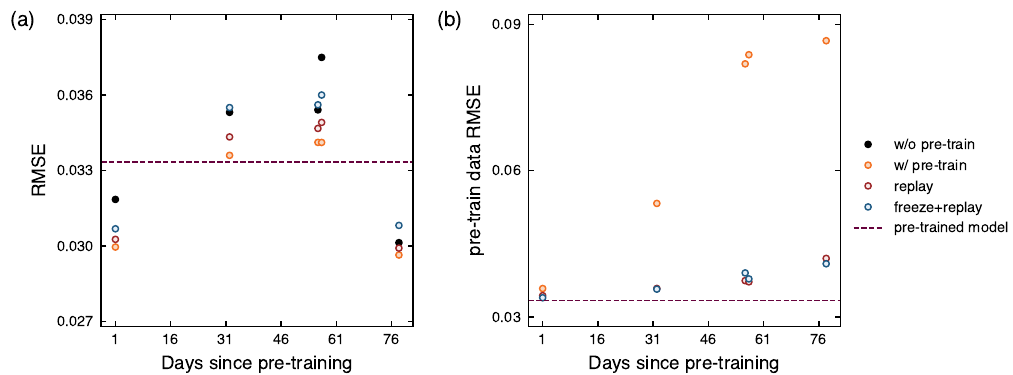}
\caption{
Performance comparison during transfer learning. 
(a) RMSE on freshly acquired data evaluated for models trained with (open orange) and without (solid black) pre-trained weights, using replay (open red) and freezing with replay (open blue) methods.
(b) RMSE evaluated on the pre-training dataset, illustrating how much the model forgets on previously learned information.
When the model was trained without any pre-training (solid black), the parameters were initialized using He initialization only on the first day. 
For all other solid black and open data points, training begins with the most recently obtained model weights, either from pre-training or a previous round of post-training.
The dashed lines indicate the average RMSE on the entire pre-training dataset of the pre-trained model.
}
\end{figure*}

\begin{figure*}
\includegraphics{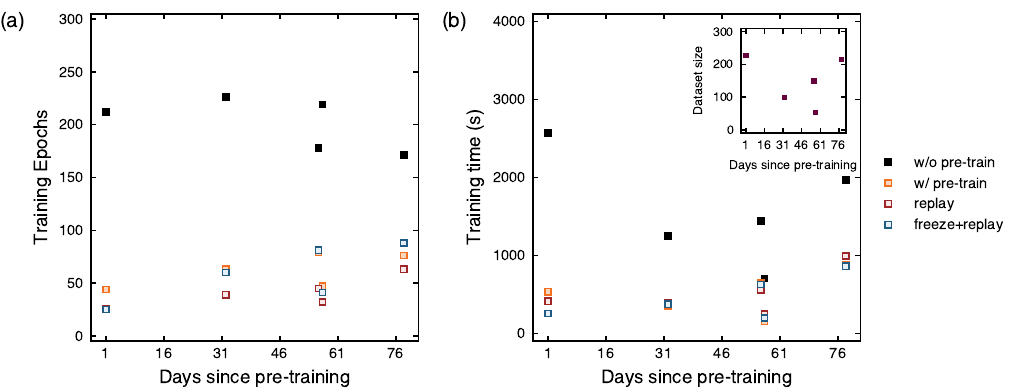}
\caption{
Training efficiency comparison during transfer learning. 
(a) Number of training epochs required until convergence for models trained with (open orange) and without (solid black) pre-trained weights, using replay (open red) and freezing with replay (open blue) methods.
(b) Total training time required until convergence for each training method.
The inset shows the number of images contained in each dataset used for post-training.
The GPU used for training was NVIDIA RTX 4090.
As in Fig. 4, the parameters were initialized using He initialization only on the first day for the model trained without any pre-training (solid black). 
For all other solid black and open data points, training began with the most recently obtained model weights, either from pre-training or a previous round of post-training.
}
\end{figure*}

The diverse dataset used for reference image reconstruction highlights the versatility of DNNs in fringe removal.
However, drifting experimental conditions inhibit direct application of pre-trained DNNs to freshly acquired data.
In Fig. 3, we display the evolution of the RMSE value for the dataset newly obtained after the model was trained, which illustrates the deterioration of model performance in reconstructing ideal reference images as the experimental system continues to drift. 
The longer the time interval between pre-training and application, the more pronounced this degradation becomes, as the model parameters gradually lose relevance under evolving experimental conditions.
Therefore, for daily operations, the network needs to be updated timely and also efficiently.

To address this issue, we employ transfer learning to adapt the pre-trained network to new data. 
By transferring knowledge from pre-trained DNNs, in general, we may arrive at optimal model weights for freshly acquired data with a shorter training time \cite{yosinski2014transferable}.
We tested three different methods to apply transfer learning in the fringe-removal algorithm.
For all the different methods, we end the training sequence with a patience value of 20 epochs.
The LRs are set to $10^{-4}$ as in the initial training. 
Refer to Figs. 8 and 9 in the Appendix for the effect of lower LR during transfer learning.

First, we start post-training on freshly acquired data with pre-trained weights.
In Fig. 4(a) we represent the average RMSE on freshly learned data with open orange circles. 
For comparison, we also show the performance of models that were individually trained on freshly acquired data in the absence of pre-training (solid black circles).
Within the training protocol and evaluation criterion, pre-trained models slightly outperform models that have not undergone pre-training.
This discrepancy would become more significant if the size of the freshly acquired data is smaller.
In the current study, the size of the post-training dataset for a single day ranges from 54 to 264 images.
Furthermore, the required training time is much shorter for pre-trained models, as displayed in Fig. 5.
The number of epochs required for training was reduced by more than 5 with pre-trained models within 30 days after the pre-training.
This shows that continuously updating on pre-trained weights would be a time-efficient method to obtain optimal model weights.

One drawback encountered in post-training on a new dataset is that the updated model is likely to forget previously learned information, even if the post-training starts from pre-trained weights.
To characterize how much the updated models forget about previously learned information, in Fig. 4(b), we show the performance of updated models on the pre-training dataset.
It is clear that the performance drastically deteriorates as the models are updated [open orange circles in Fig.~4(b)].

To allow the model not to forget previously learned information, we perform post-training of the model with a ``replay'' dataset added on top of fresh data, starting from pre-trained weights.
To this end, we randomly keep 10 images from each date in the pre-training dataset as a replay buffer.
The 10 randomly chosen images from each date in the pre-training dataset (which sum up to a total of 90 images) are combined with fresh data for post-training.
After a single round of transfer learning, we discard the 10 images from the oldest date in the replay buffer and append randomly chosen 10 images from the latest fresh data used for training.
This keeps the size of the replay buffer at a constant value of 90 images throughout the post-training process.

In Fig. 4(b), we show together the performance of updated models when training with a replay buffer (open red circles).
In contrast to the case without a replay buffer (open orange circles), the performance on pre-training data only gradually decays upon further training.
Also, both the performance on freshly acquired data and the training time are comparable to using pre-trained weights without any replay [Figs. 4(a) and 5], which indicates no detrimental effect of the replay buffer on updating efficiency.

Finally, we partially freeze the feedforward network during post-training and analyze its effect on the performance of the updated model.
Here, we freeze the entire encoding network, the bottleneck, and 6 convolutional layers in the initial upsampling layers.
The motivation for partially freezing the layers is to prevent overfitting, when the size of the indiviual post-training datasets are small compared to the entire pre-training dataset \cite{yosinski2014transferable}.
If the model starts to overfit on specific tasks, previously learned representations would be erased.
Here we employ an identical replay buffer scheme as before.
Within the training protocol and evaluation criterion, we find that the effect of the specified freeze-out does not significantly degrade the performance of the updated model on freshly acquired data [Fig. 4(a)] but also does not exhibit an improvement in the retention of prior knowledge [Fig. 4(b)].
Future improvements might involve introducing additional layers near the output or modifying the freezing strategy \cite{yosinski2014transferable}.

%%%%%%%%%%%%%%%%%%%%%%%%%%%%%%%%%%%%%%%%%%%%%%%%%%%%%
% Summary

\section{Summary and Outlook}

We have demonstrated a deep learning-based approach that reduces interference fringes from absorption signals in a dual-species atomic system.
The flexibility and comprehensive nature of the DNN, as suggested by the different scenarios in which the model inference was applied, show that this method can be straightforwardly integrated into diverse atomic absorption imaging protocols.
In addition, we have demonstrated an online model update method based on a transfer learning scheme, which is practically important for daily operations in drifting experimental environment.
One interesting extension would be to use synthetic data in training the model. 
Recently, synthetic data were used in PCA to suppress fringes with only a small set of experimentally acquired basis images \cite{song2020effective}.
Using synthetically generated data -- via translation, rotation, scaling, and shearing -- it may be possible to further improve the model without additional data acquisition.

%%%%%%%%%%%%%%%%%%%%%%%%%%%%%%%%%%%%%%%%%%%%%%%%%%%%%
% Acknowledgements

\section{Acknowledgements}
We thank Wan Zo and Myeonghyeon Kim for helpful discussions.
This work was supported by the National Research Foundation of Korea (Grants No. RS-2023-NR077280, No. RS-2023-NR119928, and No. RS-2024-00413957).

%%%%%%%%%%%%%%%%%%%%%%%%%%%%%%%%%%%%%%%%%%%%%%%%%%%%%
% Appendix

\section{Appendix}

\subsection{DNN architecture}
\begin{figure*}
\includegraphics{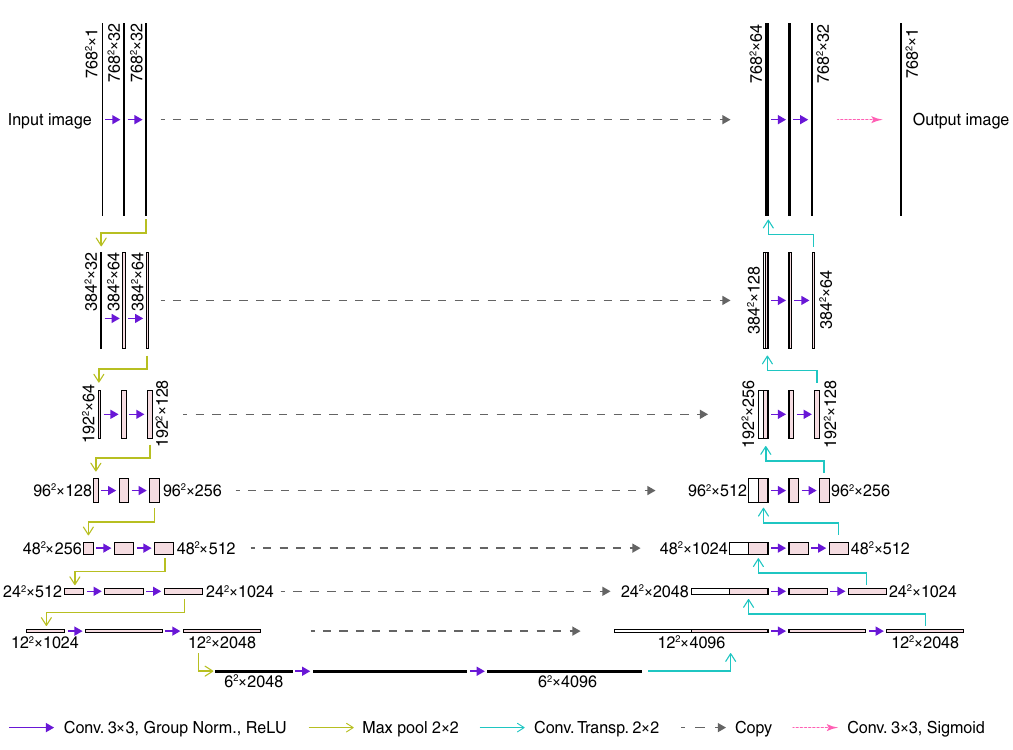}
\caption{
Concept diagram of the U-Net architecture used in this work.
The feedforward network is basically structured into an encoding stage, a bottleneck, and a decoding stage.
The skip connections between the encoding and decoding stage help recover the fine details of the input image.
}
\end{figure*}

Figure 6 depicts the pipeline of the DNN used in this work, which is inspired by the U-Net \cite{ronneberger2015u}.
The feedforward network consists of an encoding stage, a bottleneck, and a decoding stage.
The encoding stage consists of 7 max pool layers with 2 convolutional layers inserted before downsampling.
The increased depth, compared to \cite{ness2020single}, is adopted to accommodate the larger spatial resolution of the image. 
For a larger input, the network needs to scale up so that the receptive field captures the long-range contexts of the image \cite{tan2019efficientnet}.
At the bottleneck, the network exhibits the smallest spatial resolution with the highest number of feature channels.
During the decoding stage, the spatial resolution of the original input image is restored through 7 transposed convolutional layers.
For each spatial resolution, the output of each encoding layer is fed into the decoding layers so that we can recover the fine details despite the large depth \cite{he2016deep}.
In the final stage of the network, we use a sigmoid activation function to output a normalized image. 
For the freeze-out method used in transfer learning, we freeze the entire network preceding the third transposed convolutional layer. 

\subsection{Reference image completion using a DNN}

\begin{figure*}
\includegraphics{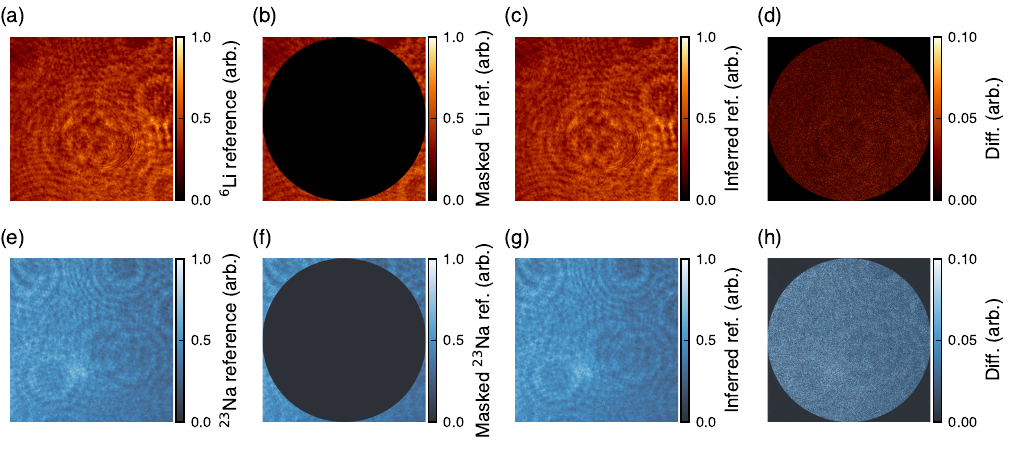}
\caption{
Examples of reference image completion by the pre-trained DNN. 
(a, e) Raw reference images for ${}^{6}$Li and ${}^{23}$Na in a standard double-shot method, respectively. 
(b, f) Masked images provided as input to the feedforward network. 
(c, g) Corresponding images reconstructed by the DNN with (b, f) used as inputs. 
(d, h) Absolute differences between the raw and reconstructed images.
Compared to (a, b, c, e, f, g), the range of the color scale has been reduced by a factor of ten for better visual contrast
}
\end{figure*}

Figure 7 demonstrates examples where the pre-trained DNN reconstructs reference images from masked inputs.
Figs. 7(a) - (d) and Figs. 7(e) - (h) show the reference image completion for the second image of a standard double-shot method in ${}^{6}$Li and ${}^{23}$Na absorption imaging, respectively.
Figs. 7(a) and 7(e) present the raw intensity profile of these reference images, while Figs. 7(b) and 7(f) display the masked images used as inputs to the DNN.
A visual comparison between Figs. 7(a) and 7(e) clearly suggests the qualitative difference between typical ${}^{6}$Li and ${}^{23}$Na reference images.
Note that the two different images were taken on the same day.

Figs. 7(c) and 7(g) illustrate the output predictions generated by the pre-trained network from the masked inputs [Figs. 7(b) and 7(f)].
The pre-trained model effectively reconstructs the missing content within the mask, even though different optical configurations were employed for probing ${}^{6}$Li and ${}^{23}$Na.
Finally, Figs. 7(d) and 7(h) depict the absolute differences between the experimentally obtained reference images [Figs. 7(a) and 7(e)] and the model-predicted reference images [Figs. 7(c) and 7(g)].

Outside the masked region, the ideal behavior of the DNN is to act as an identity mapping, thereby reproducing the input values without modification. 
In practice, we find that the average RMSE outside the circular mask is only 7.4\% of the overall RMSE observed across the entire pre-training dataset, indicating minimal contribution to the average RMSE. 
Since the masked area constitutes $\frac{\pi}{4}$ of the total image area, we isolate the model’s performance within the mask by scaling the RMSE accordingly with $\frac{4}{\pi}$. 
After applying this normalization, the average RMSE within the central masked region is found to be 0.033 for the pre-trained DNN. 
For consistency, all RMSE values reported in this work are computed after multiplying with a scaling factor of $\frac{4}{\pi}$.

\subsection{Learning rate dependence during transfer learning}

\begin{figure*}
\includegraphics{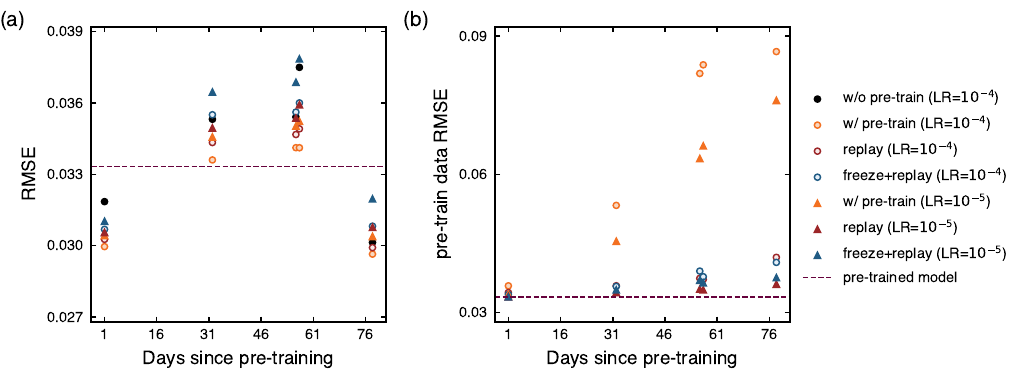}
\caption{
Learning rate (LR) dependence of network performance during transfer learning. 
(a) RMSE on freshly acquired data for LR$=10^{-4}$ and LR$=10^{-5}$ using transfer learning methods (pre-training only, replay, and freeze with replay). 
(b) RMSE evaluated on the pre-training dataset. 
As in Fig. 4, the parameters were initialized using He initialization only on the first day for the model trained without any pre-training (solid black). 
For all other solid black and open data points, training began with the most recently obtained model weights, either from pre-training or a previous round of post-training.
The dashed lines indicate the average RMSE on the entire pre-training dataset of the pre-trained model.
Circular markers are the same data points presented in Fig.~5.
}
\end{figure*}

\begin{figure*}
\includegraphics{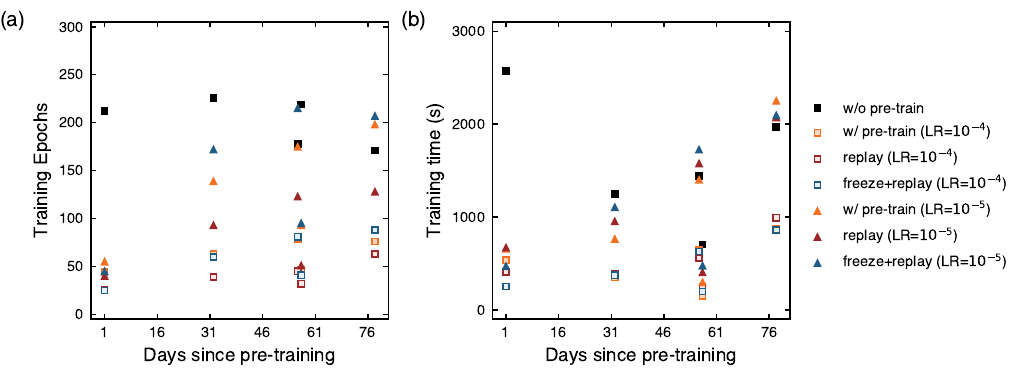}
\caption{
LR dependence of training efficiency during transfer learning. 
(a) Number of training epochs required until convergence for LR$=10^{-4}$ and LR$=10^{-5}$ under different transfer learning methods. 
(b) Total training time required until convergence.
Higher LR ($10^{-4}$) results in faster convergence, significantly reducing both training epochs and overall training time compared to LR$=10^{-5}$.
As in Fig. 4, the parameters were initialized using He initialization only on the first day for the model trained without any pre-training (solid black). 
For all other solid black and open data points, training began with the most recently obtained model weights, either from pre-training or a previous round of post-training.
Square markers are the same data points presented in Fig.~6.
}
\end{figure*}

We investigated the dependence of the network performance on the LR during transfer learning. 
Given that the nature of the reconstruction task remains relatively consistent during transfer learning, we tested a lower LR of $10^{-5}$ to explore the possibility of achieving improved model performance.
Regardless of the LR or transfer learning method, the patience value was set at 20 epochs.

Figure 8 summarizes the RMSE on (a) freshly acquired data and (b) the pre-training dataset. 
Within the training protocol and evaluation criterion, LR$=10^{-4}$ provides marginally better performance on fresh data. 
In the case of LR$=10^{-5}$, the updated model forgets less of previously learned information.
Figure 9 illustrates (a) the training epochs and (b) training time required for convergence. 
A larger LR ($10^{-4}$) leads to faster convergence compared to LR$=10^{-5}$.
The choice of learning rate should be based on the specific needs of the desired application.

%%%%%%%%%%%%%%%%%%%%%%%%%%%%%%%%%%%%%%%%%%%%%%%%%%%%%

\clearpage

\bibliography{main}

%%%%%%%%%%%%%%%%%%%%%%%%%%%%%%%%%%%%%%%%%%%%%%%%%%%%%

\end{document}